# $MgB_2$ thick film grown on Stainless steel substrate with ductility

Qing-rong Feng, Chinping Chen, Ying Lü, Zhang Jia, Jing-pu Guo,

Xiao-nan Wang, Meng Zhu, Jun Xu, and Yong-zhong Wang

Department of Physics and State Key Laboratory of Artificial Microstructure and Mesoscopic Physics, Peking University, Beijing 100871, P.R.China

## ABSTRACT

$MgB_2$ thick film was synthesized on the stainless steel substrate by the technique of hybrid physical-chemical vapor deposition (HPCVD), using Mg ingot and $B_2H_6$ as the raw materials. The film thickness is about 10 μm. The scanning electron microscope (SEM) images reveal that it is consisting of highly-densed $MgB_2$ crystals with the size ranging from 0.2 to 3 μm. The superconducting transition occurs at 38K (Tc, onset) and ends at 27(Tc, zero), giving the transition width of 11 K. The fabricated film exhibits high ductility and remains attached to the substrate after it was bent to a curvature of about 200 μm.

## 1. INTRODUCTION

Ever since the discovery of the superconductivity in $MgB_2$[1], intensive research activities on the $MgB_2$ have been stimulated both in the study of the fundamental physical properties and in the investigation of the fabrication techniques. A. Gurevich et al. has made a remark, "Due to recent advances in cryocoolers, many electric utility, fusion and high-energy physics applications may be best optimized at temperature of 10 ~ 35 K, a domain for which $MgB_2$ could provide the cheapest superconducting wires"[2]. Hence, the $MgB_2$ is a promising candidate to replace the $Nb_3Sn$ and NbTi materials in the low temperature superconducting applications. In this respect, an effective method to prepare the $MgB_2$ tape or wire is therefore important for the industrial needs.

The fabrication techniques in making the $MgB_2$ tape or wire is mostly by the method of powder in tube (PIT) or by diffusing the magnesium into the existing boron fibers. An alternative method is to grow $MgB_2$ thick film on the target substrate before making into tape or wire via appropriate engineering processes. In this paper, we report part of our ongoing project in the preparation of the MgB$_2$ thick films deposited on the stainless steel substrate. The mechanical property of the grown film with the substrate was studied as well.

## 2. EXPERIMENTAL

The setup for the HPCVD is depicted in Fig.1. The Mg ingots and the B$_2$H$_6$ gas were the raw materials. About 1 gram of the Mg ingots was placed around the stainless steel substrate at a distance of 3 mm. During the fabrication process, the background mixture, Ar (96%) + H$_2$ (4%), was introduced into the quartz glass chamber at the rate of 300 sccm (20 KPa) after the initial evacuating pumping process down to 0.1 KPa. The high frequency induction coil was then turned on to heat up the sample holder to 650 °C, and the B$_2$H$_6$ gas was subsequently introduced into the chamber at the flow rate of 30 sccm (2 KPa). The sample holder was then further heated up to 730 °C in 5 minutes, and held for additional 25 minutes to deposit the MgB$_2$ film. At the end of the process with the temperature slowly decreasing to 500 °C, the gas circuit of the B$_2$H$_6$ was cut off. Under the aforementioned conditions, the film deposition rate is estimated as 5.5 nm/s, much higher than the result reported by Pogrebnyakov et al[3]

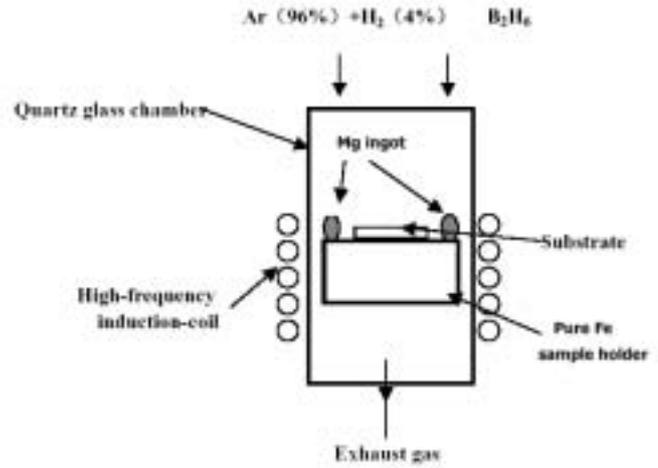

Fig.1  Experimental setup for the HPCVD.

## 3. RESULT

The XRD analysis was performed using a Philip x' pert diffractometer. The (101) peak of MgB$_2$ is the only one appearing in the spectrum without any other signal, shown in Fig. 2, indicating a well-oriented crystal growth structure and the film thickness is greater than the X-ray penetration depth, usually a few μm. The SEM observations for the morphology of the sample, shown in Fig.3, were carried out by the FEI STRATA DB235 FIB electron microscope. In Fig. 3a, the granular nature of the film is apparent. The size of these grains ranges from 0.1 to 3 μm with gaps between the grains less than 0.1 μm, similar to the thin film prepared by Zeng et al.[4]. The SEM image in the upper left inset of Fig. 3a shows that there are nanometer-sized grains existing in between the large ones. Another noteworthy feature is the mechanic property of the grown film with the substrate. Part of the film adheres to the substrate and remains intact, shown in Fig. 3b, as the sample was bent to a curvature of about 200 μm. This demonstrates that the film has the property of high tenacity, different from the bulk $MgB_2$ and some of the $MgB_2$ film reported previously[5]. The film lacerates into strips, Fig. 3c, while the sample was further bent down to a curvature of about 0.5 μm. Nonetheless, the lacerated strips still remain attached to the substrate, demonstrating a strong adhesion effect. The temperature dependence of the resistivity, ρ-T, shown in Fig.4, was performed using Quantum Design PPMS by a 4-probe technique with the electrical contacts made by cold-pressed indium. The inset is a magnified view near the Tc region. The superconducting transition occurs at 38 K, and goes to zero resistivity at 27 K, giving the corresponding transition width of 11K.

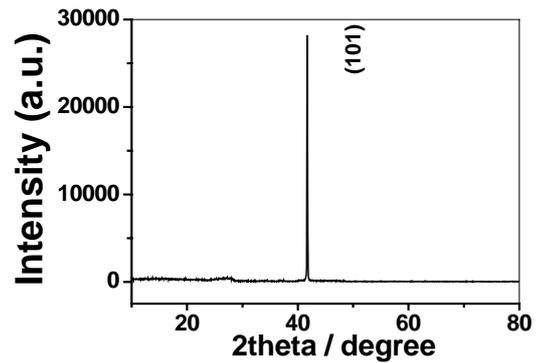

Fig. 2  The X-ray diffraction pattern of MgB$_2$ film grown on the stainless steel substrate.

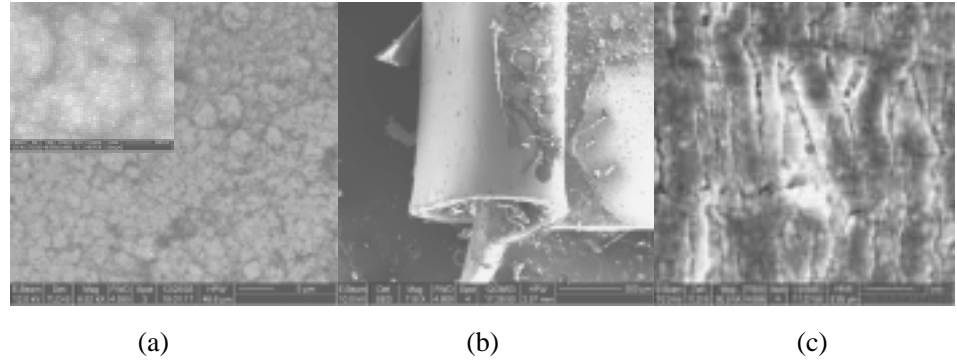

(a)          (b)          (c)

Fig.3. SEM images of $MgB_2$ film. Fig.3(a), Morphology of the film surface, showing granular nature of the film. The upper-left inset shows the image of the film observed at higher magnefication power. Fig.3 (b), Film with the substrate bending to a curvature of about 200 μm. The film remains adhering to the substrate. Fig. 3(c). Lacerated strips of the film staying attached to the substrate after bending to a curvature of 0.5 μm.

The HPCVD is an effective technique to grow thick $MgB_2$ film on the stainless substrate. The film adheres to the stainless steel substrate strongly, remaining attached to the substrate after bending down to a curvature of about 200 μm. The film also exhibits reasonable superconducting property by the ρ-T measurement. This fabrication provides an alternative method other than the PIT and the diffusion of Mg into B fiber for making the superconducting tape or wire.

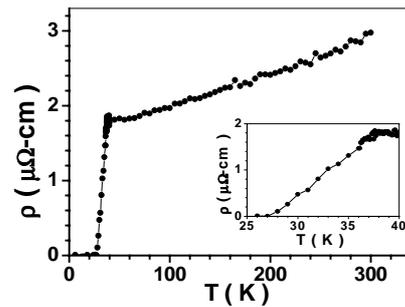

Fig.4. Temperature dependent resistivity, ρ-T. The inset shows a magnified view near the transition region. A wide superconducting transition occurs at 38 K.

## 4. ACKNOLEDGEMENT

This research was a part project of the Department of Physics of Peking University, and is supported by the Center for Research and Development of Superconductivity in China under contract No. BKBRSF-G19990646-02.